\documentstyle[preprint,aps,prd]{revtex}
\begin{document}

\tighten

\title{The Formation of Classical Defects After a Slow
\\Quantum Phase Transition}

\author{R.\ J.\
Rivers $^{1}$\thanks{r.rivers@ic.ac.uk, Permanent address:
Imperial College, London SW7 2BZ}, F. \ C.\ Lombardo$^{2}$
\thanks{lombardo@df.uba.ar}, and F. \ D.\
Mazzitelli$^{2}$\thanks{fmazzi@df.uba.ar}}
\address{{\it
$^1$ Centre of Theoretical Physics, University of Sussex, \\
Brighton, BN1 9QJ\\
$^2$ Departamento de F\'\i sica, Facultad de Ciencias Exactas y Naturales\\
Universidad de Buenos Aires - Ciudad Universitaria,
Pabell\' on I\\
1428 Buenos Aires, Argentina}}

\maketitle

\begin{abstract}

Classical defects (monopoles, vortices, etc.) are a characteristic
consequence of many phase transitions of quantum fields. We show a
model in which the onset of classical probability distributions,
for the long-wavelength modes at early times, allows the
identification of line-zeroes of the field with vortex separation.
We obtain a refined version of Kibble's causal results for defect
separation, but from a completely different approach. It is
apparent that vortices are not created from thermal fluctuations
in the Ginzburg regime.

\end{abstract}

\vskip2pc \pacs{Pacs: 03.70.+k, 05.70.Fh, 03.65.Yz}

Because phase transitions take place in a finite time, causality
requires correlation lengths to remain finite. As a result, scalar
order parameter fields $\phi ({\bf x})$ become frustrated, and
topological defects arise so as to reconcile field phases between
different correlation volumes \cite{kibble1,zurek1}. For example,
the breaking of the GUT symmetry in the early universe gives rise
to monopoles, the most familiar defect. However, they are but one
of many possibilities, that include cosmic strings (vortices),
which may be the source of high energy cosmic rays, as well as
contributing to structure formation in the early universe.

For the weak-coupling theories of the early universe, defects have
non-perturbatively large masses, comparable to the temperature
scale $k_BT_{\rm c}$ at which the transition takes place. Thus
monopoles produced in the early universe at a GUT transition might
be expected to have a mass $m\sim 10^{16}GeV$. For this reason
alone, they and other defects are manifestly classical, even
though the phase transition that produced them is intrinsically
quantum mechanical. In this letter we show how  the creation of
classical defects comes about for simple temperature quenches with
a {\it finite} quench rate through the critical temperature
$T_{\rm c}$. [The case of an instantaneous quench has been
considered elsewhere\cite{fernando2}]. Finite quench rates have
been studied by Kibble\cite{kibble2}, using causality bounds, from
which he has derived simple scaling laws that incorporate
mean-field dimensional analysis.

Although the equilibrium correlation length $\xi_{\rm eq}(t)$
diverges at the transition, the true correlation length $\xi (t)$
does not. Thus, causality imposes a maximum rate at which the
correlation length can grow and hence a maximum correlation
length, $\bar{\xi}$, at the onset of the transition. At the same
time, causality imposes a horizon outside which the fields are
uncorrelated. If we simply relate  the correlation length
$\bar{\xi}$ and the separation length between vortices
$\bar{\xi}_{\rm def}$ by $\bar{\xi}= O(\bar{\xi}_{\rm def})$, then
the density of defects is bounded, and calculable, at their moment
of formation.

If, in the vicinity of $T_{\rm c}$, $dT/dt = -T_{\rm c}/\tau_{\rm
Q}$, then dimensional analysis suggests that the earliest possible
time at which defects could be formed after the onset of the
transition is $t = O(t_{\rm K})$, where $t_{\rm K} = (\tau_{\rm
Q}/\mu^2)^{1/3}$.  This is deduced from $\dot\xi_{\rm eq}(t=t_{\rm
K})=-1$, where $\xi_{\rm eq}(t) = \vert m^{-1}(t)\vert = \mu^{-1}
\sqrt{\tau_{\rm Q}/t}$, ($\mu$ is the particle mass). We shall use
this as a benchmark in our calculations. Therefore, from causality
arguments, we derive the bound\cite{kibble1}
\begin{equation}\bar{\xi}_{\rm K}=\bar{\xi}
(t=t_{\rm K}) = \bar{\xi}_{\rm eq}\approx \mu^{-1}(\tau_{\rm Q}\mu
)^{1/3},\label{causality}\end{equation} leading to the estimate
$\bar{n}_{\rm def} = O(\bar{\xi}^{-2}_{\rm K})$ for the defect
density at the time of their production.

But there is an alternative scenario for counting defects after
the transition is completed. Simple defects have false
ground-state or vacuum at their cores where the field vanishes.
Under suitable conditions, the separation length for defects is
more sensibly derived by counting zeroes of the field as
$\bar{\xi}_{\rm def}=O(\bar{\xi}_{\rm zero})$, where
$\bar{\xi}_{\rm zero}$ measures the separation of field zeroes.
However, in principle, $\bar{\xi}_{\rm zero}$ and $\bar{\xi}_{\rm
def}$ are different correlation lengths, since not all  zeroes are
candidate defects (because zeroes occur on all scales). One needs
to count zeroes of an appropriately coarse-grained field, in which
structure on a scale smaller than an specific classical vortex
size $\xi_0 \sim \mu^{-1}=\Lambda$, is not present. At a given
time $t^*$ the separation of zeroes is $\bar{\xi}_{\rm
zero}^\Lambda (t^*)$. Then, in order that line-zeroes can be
identified with classical defects vortex cores, and
$\bar{\xi}_{\rm zero}^\Lambda (t^*)$ with $\bar{\xi}_{\rm def}$ we
need satisfy, essentially, two important conditions\cite{finland}:
I) the separation between zeroes must be insensitive to the cutoff
scale $\Lambda^{-1}\sim\mu^{-1}$; and II) the energy in field
gradients should be commensurate with the energy in classical
defects with the same density as that of line-zeroes.

We will show how decoherence can help  to satisfy these two
conditions. For weak couplings (massive defects) our main
prediction agrees qualitatively with the causal (dimensional)
prediction for counting defects, but here from a completely
different point of view, in which classicalisation of
long-wavelength (unstable) modes of the field warrants the
identification of line-zeroes with defects at the time in which
the transition has been completed.

There are further complications according as the symmetries are
global or gauged \cite{rajantie}, but the simplest of all are
global, and it is these that we shall consider here. We have
global vortices in mind, the simplest sensible defects, in which
case the order parameter field $\Phi ({\bf x})$ is a complex
scalar $\Phi =(\phi_1 + i\phi_2)/\sqrt{2}$, with action
\begin{equation}
S [\Phi ] = \int d^4x\left\{ {1\over{2}}\partial_{\mu} \phi_{\rm
a}\partial^{\mu} \phi_{\rm a} + {1\over{2}}\mu^2 \phi_{\rm a}^2 -
{\lambda\over{4}}(\phi_{\rm a}^2)^2\right\}. \label{S}
\end{equation}
With $\mu^2>0$, the $O(2)$ symmetry is broken at the scale $|\Phi
| = \eta$, $\eta^2 = \mu^2/2\lambda$. [Generalisation to global
monopoles, or even domain walls, is straightforward.]

 Global
vortices are line defects in the field, solutions to
\begin{equation}
{\delta S [\Phi]\over{\delta \Phi}}= 0, \label{classical2}
\end{equation}
around which the field phase $\theta$ ($\Phi = {\mbox h}
e^{i\theta}$) changes by $2\pi$. Considered as tubes of `false'
vacuum, with cold thickness $O(\mu^{-1})$, they have a line of
field zeroes at their cores and, once the transition is complete,
energy per unit length $\sigma =O(\mu^2/\lambda)$ (up to
multiplicative logarithmic terms $\ln (\mu\xi_{\rm def})$).

There is no unambiguous definition of the onset of classical
behaviour, least of all for the production of classical defects.
Our approach  stresses the role of {\it classical stochastic field
equations}. They arise in two forms, in Lagrange's form as
stationary phase approximations to the evolution kernel of the
density matrix, or in Hamilton's form as the dominant path in
phase space of the evolution kernel of the Wigner functional.
Initially we shall concentrate on the former. In general, if the
system is {\it closed}, quantum interference between different
field configurations forbids us from identifying the dynamical
solutions to (\ref{classical2}) for the evolving system as
describing real defects. However, if the system is {\it open}, the
environment with which it interacts can eliminate the
interference, in which case the system can be said to have {\it
decohered}.

In practice, the systems of interest to us are open. This is true,
firstly, in the trivial sense that, in the absence of
superselection rules, our order parameter fields $\Phi ({\bf x})$
interact with everything else in the universe (termed $\chi ({\bf
x}) $), but this environment is traced over when looking at the
transition.  Further, for the continuous transitions of (\ref{S})
that will interest us here, field ordering takes place through the
exponential growth of the {\it unstable} ($k<\mu$) long-wavelength
modes
 $\Phi_<({\bf x})$ of $\Phi ({\bf x})$ . The {\it stable} ($k>\mu$)
short-wavelength modes $\Phi_>({\bf x})$  do not become classical
and behave as a further part of the environment in which the
long-wavelength modes, which constitute our `system', decohere. As
a consequence, the {\it interiors} of vortices are not classical.
Since it is the field profile in the interior of the classical
vortex that carries the non-perturbatively large energy or
tension, this might be thought to be damaging, but we will see
that it is not the case. In particular, topological charge can be
classical. As a result, non-classical cores are irrelevant as far
as counting defects is concerned.

Our classical equations are, most simply, stationary phase
approximations to the evolution of the {\it reduced} density
matrix $\rho_{{\rm r}}[\phi^+_<,\phi^-_<, t]=\langle\phi^+_<\vert
{\hat\rho}_r (t)\vert \phi^-_< \rangle$, obtained by tracing out
the environments $\chi$ and $\phi_>$ (we are using $\phi_>$ to
denote the real Cartesian doublet) . In a field basis (driven by
our need to describe defects through field equations) $\rho_{{\rm
r}}[\phi^+_{<}, \phi^-_{<},t]$ evolves in time by means of the
coarse-grained effective action (CGEA) $A[\phi^+_<,\phi^-_<]$, as
we have shown in Refs.\cite{lmr,fernando1} (and in a more detailed
version in \cite{lmr2}). The CGEA takes into account the effect of
the environment on the dynamics of the system. We regard the
imaginary part of $A$ as coming from a {\it classical} noise
source $\xi (x)$ (or several such sources), say. If its
probability distribution is $p[\xi ]$, then the CGEA can be
rewritten as\cite{fernando1,fernando3}
\begin{equation}A[\phi^+_<,\phi^-_<]=-{1\over{i}} \ln  \int {\cal D}
\xi\, p[\xi] \exp\bigg\{i S_{\rm eff}[\phi^+_<,\phi^-_<,
\xi]\bigg\},
\end{equation}
where
\begin{equation}
S_{\rm eff}[\phi^+_<,\phi^-_<,\xi]= {\rm Re} A[\phi^+_<,\phi^-_<]-
\int d^4x\Big[\Delta (x) \xi (x) \Big],
\end{equation}
and $\Delta (x)$ depends on the way\cite{fernando1,GR,GM} in which
the environment couples to the $\phi_<$ field. For the simplest
case of biquadratic couplings (e.g. $\phi_<^2\chi^2$ or the
inevitable $\phi_<^2\phi_>^2$), we find $\Delta =
(\vert\Phi^+_<\vert^2 - \vert\Phi^-_<\vert^2)/2 $.

The stationary phase equation, obtained by taking the functional
variation
\begin{equation}
\left.{\delta S_{\rm eff}[\phi^+_<,\phi^-_<, \xi]\over{\delta
\phi^+_{\rm a<}}}\right\vert_{\phi^+_<=\phi^-_<} = \frac{\delta
{\rm Re}A[\phi^+_<,\phi^-_<, \xi ]}{\delta \phi^+_{\rm
a<}}\bigg|_{\phi^+_<=\phi^-_<}
-\xi\frac{\delta\Delta}{\delta\phi^+_{\rm
a<}}\bigg|_{\phi^+_<=\phi^-_<} = 0, \label{lan}
\end{equation}
is a generalised Langevin equation, with classical noise $\xi$ and
dissipation, to satisfy the fluctuation-dissipation theorem (see
Ref.\cite{lmr2} for a complete deduction of Eq.(\ref{lan})). If we
look for vortex solutions  $\Phi^{\xi}_<$ to (\ref{lan}), modelled
on the vortex solutions to (\ref{classical2}), we shall find
ourselves in some difficulty at early times, in that the noise
will cause fluctuations  (e.g. small vortex loops) that will make
the identification of a single vortex difficult.

However, the existence of classical defects depends on more than
clean vortex solutions to Eq.(\ref{lan}). The absence of quantum
interference between the stationary phase solutions
$\phi_{<}^{\xi}$ to the classical stochastic equations and their
neighbours is manifest through the increasing {\it
diagonalisation} of $\rho_{{\rm r}}[\phi^+_<,\phi^-_<, t]$. This
leads to the crucial notion of a {\it decoherence time} $t_D$,
after which $\rho_{{\rm r}}[\phi^+_<,\phi^-_<, t]$ (or, more
exactly, its real part) is effectively diagonal. The equation
(\ref{lan}) is the classical stochastic equation that we are
looking for but, as it stands, is only guaranteed to describe
classical defects {\it after} decoherence.

A digression is necessary. There is not a universal decoherence
time $t_D$. $t_D(k)$ depends on wavelength. Long-wavelength field
modes decohere first, shorter ones later, and those with $k>\mu$
never. To make the problem of diagonalisation tractable we have
restricted ourselves to field configurations $\Phi^{\pm}_<(k_0)$
peaked around  a particular wavenumber
$k_0$\cite{fernando2,lmr,fernando1,lmr2}. The order parameter
itself corresponds to $k_0=0$ but, more generally, we are
motivated by the way in which the power in the field fluctuations
is increasingly peaked around long-wavelengths in the early stage
of the transition\cite{finland,boya,Karra}. That is, if the power
spectrum $P(k,t)$ of the field fluctuations is defined by
\begin{equation}
G_<(r, t)= \langle\Phi_< ({\bf x})\Phi_<^* ({\bf 0})\rangle_t
=\int_{k<\mu} {d^3k\over{(2\pi )^3}}\,P(k,t)\, e^{i{\bf k}.{\bf
x}}
\end{equation}
then, provided the quench is fast enough, $k^2P(k,t)$ rapidly
develops a `Bragg' peak at $k^2 = k_p^2(t)<\mu^2$.  A sufficient
condition for this to happen is that
$\mu\tau_Q\lesssim\eta/\mu$\cite{Karra}. Tighter, but less
transparent bounds can be given\cite{Karra}. We restrict ourselves
to quenches permitting such a dominant momentum peak and identify
$k_0$ with it.  The relevance of this is that it is the peak in
$k^2P(k,t)$ that sets the scale for the separation of line-zeroes
in the field $\phi$. Once $k_p^2(t)\ll\mu^2$ it does not matter
where the boundary $\Lambda = O(\mu )$ between system and
environment is set. Since defects can be characterised by the
line-zeroes at their cores\cite{al}, it is the dominant wavenumber
$k_0$ that, after the onset of classical behaviour, sets the
density of the vortices that we wish to determine. This peak at
$k_0$ also warrants the fulfilment of condition (I).

This peaking stops once the field has sampled the ground state
values at $|\Phi | = \eta$. It does this by the spinodal time
$t_{\rm sp}$, defined as that value of $t$ (after the critical
temperature $T_{\rm c}$ has been crossed) for which $\langle
|\Phi_<|^2\rangle_t = \eta^2$. For the critical temperature
$T_{\rm c}$ traversed at a finite rate $\tau_{\rm Q}^{-1}$
($\mu\tau_{\rm Q}\gg 1$) we find\cite{finland,Karra,bowick}
\begin{equation}
\exp\left\{\frac{4}{3}\left({t_{\rm sp}\over{t_{\rm
K}}}\right)^{3/2}\right\} \sim {\mu\eta^2t_{\rm K}^2\over{T_{\rm
c}}}. \label{slowt}
\end{equation}
In Eq.(\ref{slowt}) the time $t_{\rm K}$ is the same Kibble causal
time introduced earlier (before Eq.(\ref{causality})). Because of
the exponential nature of the relationships, in practice $t_{\rm
sp} = O(t_{\rm K})$, but larger than it.

For $t\lesssim t_{\rm sp}$, perturbation theory can be used to
calculate $t_D$. Decoherence at scale $k_0$ occurs when the 
non-diagonal elements
of the reduced density matrix are much smaller than the diagonal
ones. In \cite{lmr2} we have evaluated $t_D$ for the longest-wavelength
modes, and found that $t_D(k_0 = 0)<t_{\rm sp}$ for all
sufficiently fast $\tau_{\rm Q}$  for a realistic class of
environments (for the {\it instantaneous} quench see
Refs.\cite{lmr,fernando1}). We have considered 
an explicit environment consisting of a large number $N$ of weakly
coupled $\chi$ fields, with coupling constants $g_i$ satisfying
$\sum_i g^2_i = O(\lambda^2)$, which permit a consistent loop
expansion, with $\mu\ll T_{\rm c}\ll\eta$, and we continue with this
hierarchy here. The expressions are complicated, even
in one-loop, and the reader is referred there for details. The
main result for the decoherence time in a slow quench (and
$k_0\approx 0$) is

\begin{equation}
\exp{\left\{ {4\over{3}}\left({t_D\over{t_{\rm
K}}}\right)^{3\over{2}} \right\}} \sim {\eta^2\over{T_{\rm c}^2}}
t_{\rm K} \sqrt{\mu T_{\rm c}}. \label{td}\end{equation}

Nevertheless, what interests us here is the shortest-wavelengths
that will have decohered by time $t_{\rm sp}$. Provided $k_0$ is
not too close to $\mu$ it is sufficient to replace $\mu t_D$ by
$t_D(k_0)\sqrt{\mu^2-k_0^2}$ in the calculations above to
determine the bound on the decoherence time for the mode $k_0$. An
immediate question is to ask what is the maximum value $k_{\rm
max}$ for which $t_{\rm D}(k_{\rm max})= t_{\rm sp}$. After this
replacement in Eqs.(\ref{td}) and (\ref{slowt}) we obtain
\[
\frac{k_{\rm max}^2}{\mu^2} \approx \ln\left({T_{\rm c}\tau_{\rm
Q}^{2/3} \over{\mu^{1/3}}}\right).
\]

Alternatively, there is a time $t(k_p)$ when the mode $k_p$ would
become the dominant wavenumber. This `quantum time' is determined
by the instabilities of the long-wavelength modes, which
increasingly bunch about the wavenumber $k^2_p(t)\sim \mu /
t(k_p)$\cite{Karra}. At time $t_D(k_{\rm max}) = t_{\rm sp}$
\begin{equation}
\frac{k^2_p(t_{\rm sp})}{\mu^2} \sim \frac{1}{\mu t_D}\sim
\frac{1}{\mu t_{\rm K}}\ln\left(\frac{T_{\rm c}}{\mu\eta^2t^2_{\rm
K}}\right).
\end{equation}
Thus, the dominant modes at time $t_{\rm sp}$ have already
decohered by this time.

Therefore, from the two previous equations we can see that

\begin{equation}\frac{k_{\rm max}^2}{k^2_p} \approx 2 \mu t_{\rm K}
\ln\left({T_{\rm c}\tau_{\rm Q}^{2/3}\over{\mu^{1/3}}}\right) > 1.
\end{equation}

That is, the modes that determine defect separation are well
decohered by time $t_{\rm sp}$. However, $t_D(k)\rightarrow\infty$
as $k\rightarrow \mu^-$ and the modes that characterise the
classical profiles of individual defects ($k\gtrsim\mu$) are not
decohered.

This makes constructing exemplary solutions to Eq.(\ref{lan}), and
averaging them, a suspect activity. Rather, we examine the second
strand of classicality, that of classical probabilities, as
expressed through {\it classical correlations}. This is understood
as the quantum system, in full (field) phase-space, mimicking the
classical dynamics. The comparison is made through the reduced
Wigner functional, defined as
\begin{equation}
W[\phi_{<},\pi_{<};t] = \int{\cal D} \eta_{<}
~~e^{i\int\pi_{<}\eta_{<}} ~ \langle\phi_{<}- \eta_{<}|\rho_{\rm
r}(t)| \phi_{<}+ \eta_{<}\rangle.
\end{equation}
As is well known, $W[\phi_{<},\pi_{<};t]$ itself is not guaranteed
to be positive until decoherence has occurred \cite{diana},
although
\begin{equation}
P[\phi_{<}]_t = \int {\cal D}\pi_{<}\,W[\phi_{<},\pi_{<};t]
=\langle\phi_{<}|\rho_{\rm r}(t)|\phi_{<}\rangle >0, \label{p}
\end{equation}
is always a true probability distribution for the field
configuration $\phi_<$ at time $t$.

After decoherence, when $W[\phi_{<},\pi_{<};t]$ is
positive, at least for long-wavelength modes, it can be identified
with the Fokker-Planck probability distribution function $P^{\rm
FP}_t[\phi_{<},\pi_{<}]$, and $P[\phi_{<}]_t$ can be equally
identified as the Fokker-Planck probability $P^{\rm
FP}[\phi_{<}]_t$. Since the Fokker-Planck equation is the obverse
of the Langevin equation, at this time we can equally average over
Langevin noise $\xi$ or over $P^{\rm FP}[\phi_{<}]_t$.

However, as we observed earlier, the validity of the classical
stochastic equations is, of itself, not sufficient to guarantee
classical defects with energy profiles derived from
(\ref{classical2}). We do not have defects until the fluctuations
around them are negligible, and this is associated with the
peaking of the power spectrum (and the validity of our
approximation of adopting a dominant wavenumber) rather than
decoherence or classical correlations directly\cite{finland,RKK}.
To see this, suppose we can calculate $P[\phi_{<}]_t$ for all $t$.
This permits us to calculate the equal time n-point correlation
functions, in which we restore the subscripts $a,b,... = {1,2}$,
\begin{equation}
G_{<\rm ab..c}^{(\rm n)}({\bf x_1,..,x_{\rm n}},t)= \int{\cal
D}\phi_{<} P_t[\phi_{<}]\phi_{<\rm a} ({\bf x_1})...\phi_{<\rm
c}({\bf x_{\rm n}})\label{n}
\end{equation}
for all times $t<t_{\rm sp}$.  As we know, equal-time correlators
are all we need to calculate densities of field zeroes
(line-zeroes, etc.). However, if all we are going to use is
$P[\phi_{<}]_t$, the diagonal matrix element of $\rho_{\rm r}(t)$,
there is no real need to construct the Wigner functional. We can
just do a calculation of $P[\phi_{<}]_t$ from the start, along the
lines of Ref.\cite{boya}. The weak interaction with the
environment that is so successful in diagonalising the density
matrix has only a small effect on the diagonal matrix elements
$P[\phi_{<}]_t$ for most of the time until $t_{\rm sp}$.

The major characteristic of defects is their topological charge.
We now see how localised topological charge precedes the
appearance of vortices, and gives an estimate of their density
when they do appear. Let us now consider the line-zero ensemble density 
$n_{\rm zero}(t)$ and the line-zero separation $\xi_{\rm zero}(t)$  for
the long-wavelength mode fields. When the Gaussian
approximation\cite{halperin,maz} is satisfied (as happens for
$t<t_{sp}$) they are determined completely by the {\it
short-distance} behaviour of $G_<(r, t)$ as
\begin{equation}
n_{\rm zero}(t) = \frac{1}{2\pi\xi^2_{\rm zero}(t)} =
\frac{-1}{2\pi}\frac{G''_<(0, t)}{G_<(0, t )}. \label{ndeff}
\end{equation}
Since $G_< (r,t)$ has short-wavelength modes on the scale of a
classical vortex  removed, $G_<(0,t)$ is finite. Initially, with
$P(k,t)$ large for $k\gtrsim\mu$, $G_<(r,t)$ is very dependent on
the value of the cut-off $\Lambda$. As a result, line-zeroes are
extremely fractal, with a separation $\xi_{\rm zero}(t)$
proportional to the scale at which they are viewed, and are
certainly not candidates for defects. Only once the peak at
$k=k_0$ is firmly in the interval $k<\mu$ does $\xi_{\rm zero}(t)$
becomes insensitive to a cut-off $O(\mu^{-1})$. This means that
line-zeroes are straight at this scale, although they can be
approximately random walks at much larger scales (condition (I)).

Further, for this early period, when the self-consistent
mean-field approximation is valid, the field energy $\langle
E\rangle_t$ of the system field $\phi_<$ in a box of volume ${\cal
V}$ becomes\cite{finland,RKK}
\begin{eqnarray}
\langle E\rangle_t &=& {\cal V}[\langle |\nabla\phi_<|^2\rangle_t
+\lambda(\eta^2 - \langle |\phi_<|^2\rangle_t)^2]\nonumber
\\ &=&
{\cal V}[2\pi n_{\rm zero}(t) G_<(0,t)+ \lambda(\eta^2 -
G_<(0,t))^2]\label{E2}
\\
 &=& 2\pi L_{\rm zero}(t)G_<(0,t) +{\cal V}[\lambda(\eta^2
- G_<(0,t))^2]. \label{E3}
\end{eqnarray}
As before, $\chi$-field fluctuations are absorbed in the
definition of $\mu^2 = \lambda\eta^2$. Eq.(\ref{E2}) is obtained
by using (\ref{ndeff}) and $L_{\rm zero}(t)={\cal V} n_{\rm zero}$
is the total length of line-zeroes, on a scale $\mu^{-1}$, in the
box of volume ${\cal V}$.

We understand Eq.(\ref{E3}), valid from time $t=0$, when $G(0,0) =
O(\mu^2)$, until time $t\approx t_{\rm sp}$, when
$G(0,t_{sp})\simeq \eta^2$, as follows. At early times most of the
system field energy (proportional to ${\cal V}$) is in
fluctuations not associated with line zeroes, arising from the
field potential. As time passes their energy density decreases as
the system field approaches its post-transition value, becoming
approximately zero. In addition there is a term, arising from the
field gradients, proportional to the length $L_{\rm zero}$ of
line-zeroes, whose energy per unit length increases from
$O(\mu^2)$ to $\eta^2$. At time $t_{\rm sp}$, when the fluctuation
energy can be ignored, we have
\begin{equation}
\langle E\rangle_t \sim L_{\rm zero}\sigma, \label{E}
\end{equation}
essentially the energy required to produce a {\it classical}
vortex tangle of length $L_{\rm zero}$ (up to $O(1)$ factors from
the logarithmic tails). Eq.(\ref{E}) completes condition (II) in
order to relate line-zeroes with separation length between
defects. Although these line zeroes have the topological charge
and the energy of classical vortices, they are not yet
fully-fledged defects.  However, by the end of the linear regime,
when decoherence on the scale $\xi_{\rm zero}(t)$ has been
effected, the final coupling of radial to angular modes that turns
these proto-vortices into vortices incurs no significant energy
change, and $n_{\rm zero}$ of Eq.(\ref{ndeff}) is a reliable guide
for the initial vortex density. We stress that, as far as counting
vortices is concerned, all that matters is how the power in the
field fluctuations is distributed. The distance between defects is
the relevant wavelength, and not the defect size, which shows no
decoherence.

This distance can be computed from the two point function of the
field (see Eq.(\ref{ndeff})).  {\it Up to logarithms} in
$\tau_{\rm Q}$ and the other parameters of the theory it is given
by \cite{finland,Karra,RKK},
\begin{equation}
\xi_{\rm zero}(t\approx t_{\rm sp}) = {\bar\xi}_{\rm zero} \approx
\mu^{-1}(\tau_{\rm Q}\mu)^{1/3}, \label{kcaus}
\end{equation}
in accordance with the predictions of Kibble\cite{kibble2}, but
for very different reasons. Kibble correctly argues that the
long-distance adiabatic correlation length $\xi_{\rm eq}$ (that
would diverge at $T=T_{\rm c}$) freezes in at time $t_{\rm K}$
{\it before} the transition with value ${\bar\xi}$, and this sets
the scale for defect separation. However, in our picture defects
do not form until time $t_{\rm sp}$ {\it after} the transition
with a separation given by that of line-zeroes, that is
arbitrarily small at time $t_{\rm K}$, and at the transition
itself. That this is directly related to ${\bar\xi}$ is a
consequence of dimensional analysis. This constraint was used to
bound the production of monopoles\cite{kibble3} and cosmic
strings\cite{joao} (vortices) in the early universe. [If we had
broken an $O(3)$ symmetry (${\rm a}=1,2,3$) the corresponding
defects would be global monopoles, giving qualitatively similar
conclusions.] This is all for flat space-time.

In summary, the mechanism for the production of classical vortices
that we have proposed here has several parts. Firstly, the
environment renders the long-wavelength modes, wavenumber $k_0$,
of the order-parameter field classical at early times $t_D (k_0)$,
by or before the transition is complete at time $t_{\rm sp}$. In
particular,  those on the scale of the separation of the
line-zeroes that will characterise the classical defects will have
decohered, even though the field modes on the scale of classical
vortex thickness do not decohere. For all that, the field
possesses classical correlations at early times by virtue of the
quasi-Gaussian nature of the regime. For the longer-wavelength
modes the classical behaviour of the field is expressed through
the classical Langevin stochastic equations that it satisfies.
However, for line-zeroes to mature into vortex cores classical
stochastic equations (\ref{lan}) or, equivalently, classical
Fokker-Planck equations are not enough. The field needs to have an
energy profile commensurate with the vortex solutions to the
ordinary classical Euler-Lagrange equations (\ref{classical2}).
This requires that the field fluctuations are peaked around
long-wavelengths, to avoid fluctuations causing wiggles in the
cores and creating small cortex loops, a related condition
satisfied in our models. The resultant density of line-zeroes can
already be inferred in the linear regime, whose topological
charges are well-defined even though close inspection of their
interior structure is not permitted classically.

The final result for their density, at the time $t_{\rm sp}$ of
their formation is, up to logarithms, that proposed by Kibble
\cite{kibble2} from dimensional analysis of mean-field theory,
although the reasoning is very different. Although logarithms
introduce different scales in principle, in practice they have no
qualitative effect.  It is apparent that vortices are not created
from thermal fluctuations in the Ginzburg regime, as suggested in
the first instance by Kibble in an earlier paper\cite{kibble1}.

\acknowledgments F.C.L. and F.D.M. were supported by Universidad
de Buenos Aires, CONICET (Argentina), Fundaci\'on Antorchas and
ANPCyT. R.J.R. would like to thank the European Science Foundation
for support through its COSLAB programme, and the Rockefeller
Foundation at Bellagio for hospitality, where this work was
completed.

\end{document}